\documentclass[aps,prd,eqsecnum,11pt,showpacs,superscriptaddress,nofootinbib]{revtex4-2}
\usepackage{geometry}                		
\usepackage{graphicx}				
\usepackage{amssymb}
\usepackage{graphicx}
\usepackage{amssymb,amsmath}
\usepackage{amsfonts,mathrsfs}

\newcommand{\be}{\begin{equation}}
\newcommand{\ee}{\end{equation}}
\newcommand{\bea}{\begin{eqnarray}}
\newcommand{\eea}{\end{eqnarray}}
\newcommand{\bes}{\begin{subequations}}
\newcommand{\ees}{\end{subequations}}

\begin{document}

\title{Some considerations on 
BEC analogue black holes}

\author{Roberto~Balbinot}
\email{Roberto.Balbinot@bo.infn.it}
\affiliation{Dipartimento di Fisica dell'Universit\`a di Bologna and INFN sezione di Bologna, Via Irnerio 46, 40126 Bologna, Italy}

\author{Alessandro~Fabbri}
\email{afabbri@ific.uv.es}
\affiliation{Departamento de F\'isica Te\'orica and IFIC, Universidad de Valencia-CSIC, C. Dr. Moliner 50, 46100 Burjassot, Spain}

\begin{abstract}
In this paper, dedicated to the memory of A. Aurilia, we will review some basic features of Hawking's black hole radiation and compare them with the corresponding ones present in Bose-Einstein condensate analogue black holes.
 \end{abstract}

\maketitle

\section{Preface}

This paper is dedicated to the memory of Antonio Aurilia, a theoretical physicist with a wide range of interests. He could move easily with competence from Quantum Gravity to QCD.
One of us (R.B.) had the pleasure to work with him many years ago on a project dedicated to classical and quantum aspects of shell dynamics.
Here we give a personal account  on a field, analogue gravity,  which has tremendously grown in recent years and has allowed to verify experimentally, although in an indirect way, 
one of the most spectacular prediction of modern theoretical physics, obtained by an encounter of General Relativity and Quantum Mechanics, namely Hawking's Black Hole (BH)  emission.

\section{Hawking radiation and analogue models}
\label{sec:1}

One of the most striking and unexpected features of BHs is that, despite their name, they emit radiation. This is the astonishing discovery of S. Hawking in 1974 \cite{Hawking:1974sw}. The number of particles emitted follows a thermal law
\be \label{uno}
N_\omega=\frac{1}{e^{\frac{\hbar\omega}{k_BT_H}}-1}\ ,
\ee
where $T_H$ is the emission temperature of the hole, $k_B$ the Boltzmann's constant and $\hbar$ the reduced Planck's constant.
$T_H$ is the Hawking temperature which is proportional to the surface gravity $\kappa$ of the BH horizon
\be 
\label{due}
T_H=\frac{\hbar \kappa}{2\pi k_B}\ .
\ee
This discovery has allowed a beautiful synthesis between gravity and thermodynamics mediated by Quantum Mechanics, as can be appreciated by eq. (\ref{due}), and led to the identification of the laws of BH mechanics  \cite{bch} with the fundamental principles of thermodynamics. One can consider this synthesis as the basic support for the existence of Hawking BH radiation since a direct experimental verification seems till now quite impossible. 
Indeed, the emission temperature $T_H$ scales with the inverse of the BH mass $M$, roughly
\be 
T_H \sim 10^{-6} (\frac{M_s}{M})\ , 
\ee
where $M_s$ is the solar mass. 
Since BHs formed by stellar gravitational collapse have masses bigger than $M_s$, $T_H$ appears by far much lower than the cosmic microwave background radiation ($\simeq 2.7 K$) to allow a direct observation. Furthermore, so far, there is no evidence of an emission of this type related to a primordial population of BHs whose mass can be much lower then $M_s$ \cite{carr}. 

\noindent Beside this, also from the theoretical point of view Hawking's derivation of his famous result is not free of criticisms. The theoretical
scheme assumed by Hawking in his seminal work is that of Quantum Field Theory (QFT) in curved space-time. In this hybrid framework
matter is quantized according to QFT while gravity is treated classically according to General Relativity. The limit of validity of this approximation of a yet unknown quantum gravity theory is expected to be set by the Planck scale
\be l_P=(\frac{G\hbar}{c^3})^{1/2} \sim 10^{-33}cm\ ,
\ee
or in terms of energy $E_P\sim 10^{19}GeV$. 

\begin{figure}[h]
\centering \includegraphics[angle=0, height=2.6in] {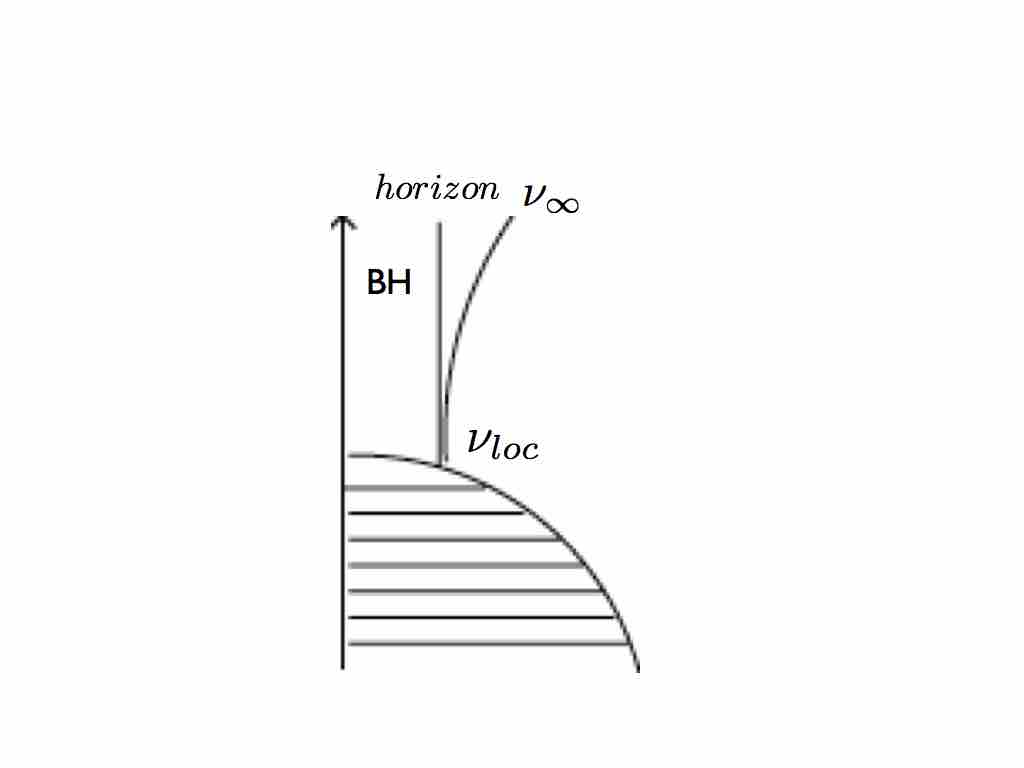}
\caption{For the Hawking particle, the local frequency close to the horizon is exponentially amplified with respect to its
frequency measured at infinity and the corresponding energy becomes transplanckian.}
\label{gravcoll}
\end{figure}

\noindent Hawking radiation can be viewed as a conversion of vacuum fluctuations near the horizon into pairs of real on shell particles, of which one escapes to infinity and constitutes Hawking radiation, the other, called ``partner'', is swallowed by the BH. Now, suppose one observes at infinity a Hawking particle of frequency $\nu_\infty$. Tracing back in time the particle's trajectory till the horizon region where it originates (see Fig. \ref{gravcoll}) one finds that there the local frequency of the particle is exponentially amplified and rapidly the corresponding energy becomes much bigger than $E_P$. 
This regime is outside the range of validity of the framework used (QFT in curved space-time). So one might wonder if Hawking's BH radiation is a real phenomenon or just an artifact of an approximation pushed far beyond its limits. This is the so called ``transplanckian problem'' \cite{ted}.

\noindent Partially motivated by this incongruence, in recent years a new branch of research has grown called `analogue gravity models'' \cite{livrevrel} which aims to reproduce, in condensed matter systems (Bose-Einstein condensates, BEC, are the most investigated) manageable in laboratory, some features of gravitational systems, most notably Hawking's BH thermal radiation. 
The analogy relies on an exact mathematical correspondence between the propagation of sound waves in the condensed matter system under the hydrodynamical approximation and the propagation of light in a gravitational field.

\noindent As shown by Unruh \cite{unruh}, the propagation of sound in an inhomogeneous fluid is governed by an effective curved space-time metric (the acoustic metric) which, if the flow turns from subsonic to supersonic, has exactly a BH form. The transition occurs at the corresponding horizon called ``sonic horizon''. Basically, sound waves are trapped and dragged by the supersonic flow as light inside a BH. So, once the sound waves are quantized one can envisage the occurrence of Hawking like radiation of phonons in a fluid which turns supersonic. 

\noindent A problem similar to the transplanckian one seems to appear also here because of the very short wavelengths of the emitted phonons near the sonic horizon breaking down the hydrodynamical (large wavelength) approximation. However for the analogue models, unlike in the gravitational case, we know exactly the short distance quantum description of the system and so we can perform an ab initio analysis based  on this fundamental description without any use of the gravitational analogy. One can then, at the end, compare the results with the predictions based on the analogy to test how this is reliable. One finds that there is a regime in which the gravitational analogy based on QFT in curved space-time does indeed predict quite well the exact result, while in other regimes the analogy fails dramatically.

\section{BEC and gravitational black holes}

\noindent Let us analyse in some detail how a supersonically flowing BEC can mimic some of the characteristic features of a BH. 

\noindent The fundamental bosonic field operator $\hat\Psi(t,\vec x)$ in a BEC can be splitted as a sum of a mean-field classical field $\Psi_0$ describing the condensate and a quantum operator describing quantum fluctuations above the condensate \cite{ps}
\be \label{cinque}
\hat \Psi (t,\vec x)= \Psi_0(t,\hat x)\left[ 1+ \hat \phi(t,\hat x) \right]\ .
\ee
$\Psi_0$ satisfies the Gross-Pitaevski 
equation  
\be \label{sei}
i\hbar \frac{\partial \Psi_0}{\partial t}=\left(-\frac{\hbar^2}{2m}\vec\nabla^2 + V_{ext}+g|\Psi_0|^2\right)\Psi_0\ ,
\ee
where $V_{ext}$ is the external potential, $m$ the mass of the atoms and $g$ the effective coupling constant.
The fluctuations quantum operator $\hat\phi$ satisfies the Bogoliubov-de Gennes (BdG) equation
\be \label{sette}
i\hbar\frac{\partial \hat\phi}{\partial t}=-\left( \frac{\hbar^2}{2m}\vec\nabla^2 +\frac{\hbar^2}{m}\frac{\vec\nabla\Psi_0}{\Psi_0}\ \vec\nabla\right)\hat\phi + gn(\hat\phi + \hat\phi^\dagger)
\ee
where $n=|\Psi_0|^2$ is the condensate density.

\begin{figure}[h]
\centering \includegraphics[angle=0, height=2.6in] {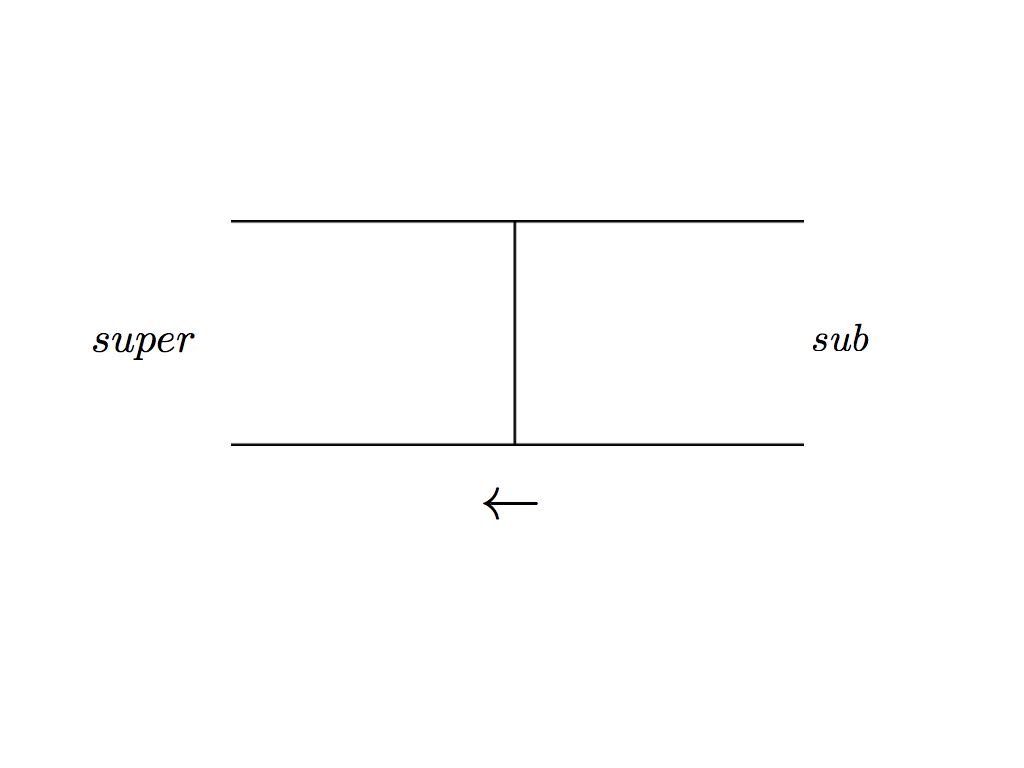}
\caption{Schematic representation of an acoustic black hole.}
\label{sub-super}
\end{figure}

\noindent Let us assume that the condensate is flowing from right to left making a transition from a subsonic flow to a supersonic one (Fig. 
\ref{sub-super}).

\begin{figure}[h]
\centering \includegraphics[angle=0, height=2.6in] {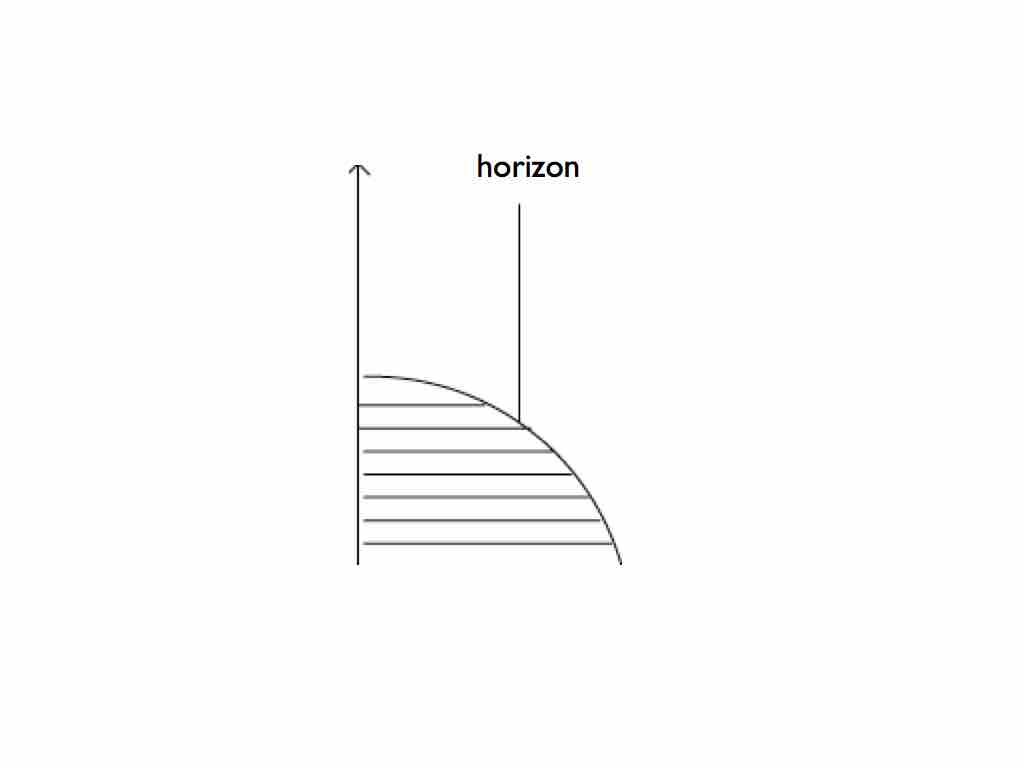}
\caption{Black hole formed by gravitational collapse.}
\label{bh-grav-coll}
\end{figure}

\noindent Consider now a BH formed by the gravitational collapse of a star (Fig. \ref{bh-grav-coll}).
The resulting gravitational field is described by the metric tensor $g_{\mu\nu}$ solution of the Einstein equations
\be \label{otto}
R_{\mu\nu}-\frac{R}{2}g_{\mu\nu}=8\pi G T_{\mu\nu} 
\ee
where $R_{\mu\nu}$ is the Ricci tensor, $R$ is the scalar curvature and $T_{\mu\nu}$ the energy momentum tensor of the matter distribution. 
In the vacuum space outside the star $T_{\mu\nu}=0$ and eqs. (\ref{otto})  reduce to $R_{\mu\nu}=0$. In this part of spacetime the solution, describing the BH, can be written in Painlev\'e-G\"ullstrand form 
\be \label{nove}
ds^2=-(c^2-V^2(x))dt^2-2V_idx^idt+dx_idx^i
\ee
where $V(x)$ is some function; for example for a Schwarzschild BH one has $V=-\sqrt{\frac{2GM}{c^2r}}$.
In this BH spacetime  consider a massless scalar quantum field $\hat\varphi$ satisfying
\be \label{dieci}
\nabla_\mu\nabla^\mu \hat\varphi =0\ ,
\ee
where the covariant derivative $\nabla_\mu$ is calculated from the metric eq. (\ref{nove}).

\noindent What do these two systems, one, the BEC described by eqs. (\ref{sei}), (\ref{sette}), the other, the BH plus scalar field satisfying eqs. (\ref{otto}), (\ref{dieci}), have in common?

\noindent If we use a density-phase representation of the BEC bosonic field operator $\hat\Psi$ 
\cite{livrevrel} (see also \cite{bcfmr})
\be
\hat\Psi=\sqrt{n+\hat n_1}e^{i(\theta+\hat\theta_1)}\simeq \Psi_0(1+\frac{\hat n_1}{2n}+i\hat\theta_1)
\ee
the BdG equation (\ref{sette}) reduces to a pair of equations of motion for the density fluctuation $\hat n_1$ and the phase fluctuation
$\hat \theta_1$ , namely
\bea
&& \hbar \partial_t \hat\theta_1=-\hbar \vec v\ \vec\nabla\hat \theta_1- \frac{mc^2}{n}\hat n_1 + \frac{mc^2}{4n}\xi^2\vec\nabla[n\vec\nabla(\frac{\hat n_1}{n})]\ , \label{undicia} \\
&& \partial_t\hat n_1=-\vec\nabla (\vec v \hat n_1 + \frac{\hbar n}{m}\vec\nabla\hat \theta_1) \label{undicib} \ .
\eea
Here a fundamental length enters, the ``healing length'' $\xi=\frac{\hbar}{mc}$ defined in terms of the local speed of sound $c_s=\sqrt{\frac{gn}{m}}$ and $\vec v=\frac{\hbar}{m}\vec\nabla \theta$ is the local speed of the flow.
If one is probing the system on scales $\gg \xi$ (hydrodynamical approximation) the last term in eq. (\ref{undicia}) can be neglected and we can write the density fluctuation $\hat n_1$ as 
\be 
\hat n_1= -\frac{\hbar n}{mc^2}[ \vec v \vec \nabla \hat \theta_1+\partial_t \hat \theta_1 ]\ ,
\ee
which, when inserted in (\ref{undicib}), gives the equation of motion for the phase fluctuation
\be \label{tredici}
-(\partial_t+\vec \nabla\vec v)\frac{n}{mc^2}(\partial_t + \vec v \ \vec\nabla)\hat\theta_1 + \vec\nabla (\frac{n}{m}\vec\nabla\hat\theta_1)=0\ .
 \ee
Remarkably, (\ref{tredici}) can be rewritten as
\be \label{quattordici}
\bar \nabla_\mu\bar\nabla^\mu\hat\theta_1 =0\ ,
\ee
where the covariant derivative $\bar\nabla_\mu$ is calculated from the ``acoustic metric"
\be \label{quindici}
d\bar s^2=\frac{n}{mc}[-(c_s^2-v^2)dt^2-2v^idx_i dt+dx^idx_ i]
\ee
showing that the propagation of sound waves in a BEC is governed by an effective spacetime metric. 

\noindent One sees at glance the amazing similitude between eqs. (\ref{quattordici}), (\ref{quindici}) for the BEC and eqs. (\ref{nove}), (\ref{dieci}) for the massless scalar field in the BH spacetime. This is the core of the analogy: in particular, both metrics exhibit a horizon, at $|V|=c$ for the BH and $|v|=c_s$ for the BEC (the sonic horizon). The corresponding surface gravity is 
\be  \label{sedici}
\kappa=\frac{1}{2c_s}\frac{d}{dn}(c_s^2-v^2)|_{hor}\ ,
\ee
where $n$ is the normal to the horizon, and a similar expression for the BH (with $c_s\to c,\ v\to V$). 

\noindent Eqs. (\ref{nove}, \ref{dieci}) are the starting point of Hawking derivation of his famous result which is completely kinematical. So, based on the analogy one expects the supersonic BEC of Fig. (\ref{sub-super}) to radiate thermal phonons in the subsonic region at a temperature given by Eq. (\ref{due}),
which would open the possibility to reveal the Hawking effect in BEC. 

\noindent One should however proceed with care since eq. (\ref{quattordici}) is just an approximation of the BdG eqs. (\ref{undicia}), (\ref{undicib}),
an approximation in the long wavelength regime $\lambda\gg \xi$. As announced we encounter here the same consistency problem
of the transplanckian frequency in the BH case since the wave equations are the same ((\ref{quattordici}) and (\ref{dieci})) and
both predict an unbounded growth of the frequency of the modes near the horizon.
However, we have now the complete quantum microscopic theory, given by the BdG eqs. (\ref{undicia}), (\ref{undicib}) and we can 
study, starting directly from these, the presence or not of Hawking like phonon radiation in the BEC system without making any
use of the hydrodynamic approximation, i.e. of the gravitational analogy.

\noindent To simplify the exposition let us assume that the BEC flow is stationary and that asymptotically (both in the subsonic and in the supersonic regions) the condensate is homogeneous. In these regions the dispersion relation of the plane waves solutions
of the BdG eqs. (\ref{undicia}), (\ref{undicib}) take the form
\be \label{diciotto}
\omega - vk=\pm c_s\sqrt{k^2+\frac{\xi^2k^4}{4}}\ ,
\ee
where $\omega$ is the conserved frequency and $k$ the corresponding wave vector.
Neglecting the last term in eq. (\ref{diciotto}), the relation becomes linear
\be \label{diciannove}
\omega= k(v\pm c_s)\ ,
\ee
which would correspond to eq. (\ref{quattordici}) and similarly to eq. (\ref{dieci}): this is the ``relativistic'' dispersion relation.
So the hydrodynamical approximation leading  to eq. (\ref{sedici}) misses the dispersive character of wave propagation
in a BEC.


\begin{figure}[h]
\centering \includegraphics[angle=0, height=2.1in] {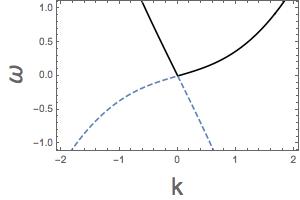}
\caption{BEC dispersion relation in the subsonic region.}
\label{bec-sub}
\end{figure}

\begin{figure}[h]
\centering \includegraphics[angle=0, height=2.1in] {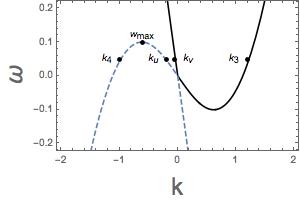}
\caption{BEC dispersion relation in the supersonic region.}
\label{bec-sup}
\end{figure}

\begin{figure}[h]
\centering \includegraphics[angle=0, height=2.8in] {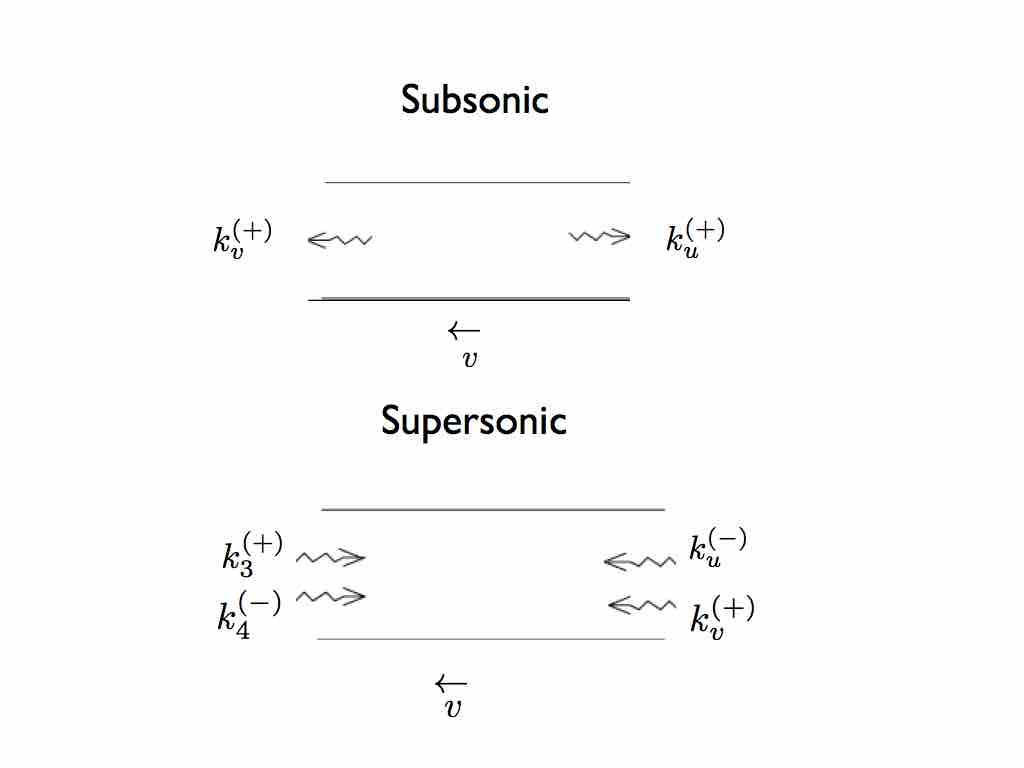}
\caption{Modes propagation in the subsonic (upper panel) and supersonic (lower panel) regions.}
\label{modes-propagation}
\end{figure}

Let us give a look at Figs. \ref{bec-sub}, \ref{bec-sup} representing, graphically, the dispersion relation (\ref{diciotto}) in, respectively, the subsonic and supersonic regions. This latter, as an equation in $k$, for fixed $\omega$, has in general four roots. In the subsonic case, two of them are real and two complex conjugates. The real ones, $k_v^{(+)}$ and $k_u^{(+)}$, correspond one to a wave propagating downstream along the flow ($k_v^{(+)}$), the other
to a wave propagating upstream ($k_u^{(+)}$). See the upper panel of Fig. (\ref{modes-propagation}).

In the hydrodynamic approximation they reduce to the two solutions of the relativistic dispersion relation (\ref{diciannove}). 
The supersonic case is much more involved. There is a threshold frequency $w_{max}$ which scales as $\xi^{-1}$. 
For $0<\omega<\omega_{max}$ there are four real roots, of which $k_v^{(+)}, k_u^{(-)}$ are similar to the ones found in the subsonic case. The novelty is that $k_u^{(-)}$ is dragged by the flow and propagates downstream like $k_v^{(+)}$. These roots belong to the linear part of the dispersion relation. $k_3^{(+)}$ and $k_4^{(-)}$ propagate upstream, notwithstanding the supersonic character of the flow. 
See the lower panel of Fig. (\ref{modes-propagation}).
So, while the propagation of the hydrodynamical modes $k_v^{(+)}, k_u^{(-)}$ can be well described by the wave equation (\ref{quattordici}) 
in the acoustic metric (\ref{quindici}), the propagation of $k_3^{(+)}$ and $k_4^{(-)}$ is completely dispersive. These modes do not feel at all the acoustic metric and the sonic horizon. Above $\omega_{max}$ only two solutions remain real. 

\begin{figure}[h]
\centering \includegraphics[angle=0, height=2.1in] {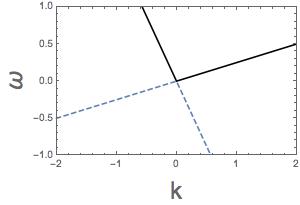}
\caption{Linear dispersion relation in the subsonic region.}
\label{linear-sub}
\end{figure}

\begin{figure}[h]
\centering \includegraphics[angle=0, height=2.1in] {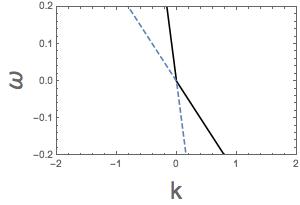}
\caption{Linear dispersion relation in the supersonic region.}
\label{linear-sup}
\end{figure}

\noindent For comparison, the relativistic dispersion relations are plotted in Figs. \ref{linear-sub}, \ref{linear-sup}. 

\noindent The dashed part in the plots of Figs. \ref{bec-sub}, \ref{bec-sup}, \ref{linear-sub}, \ref{linear-sup} represents modes with negative norm. One sees that - in the supersonic region - $k_u^{(-)}$ and $k_4^{(-)}$ have negative norm (this explains the superscript ${}^{(-)}$)
and positive frequency $\omega$ or, equivalently, positive norm and negative frequency. 

\begin{figure}[h]
\centering \includegraphics[angle=0, height=2.4in] {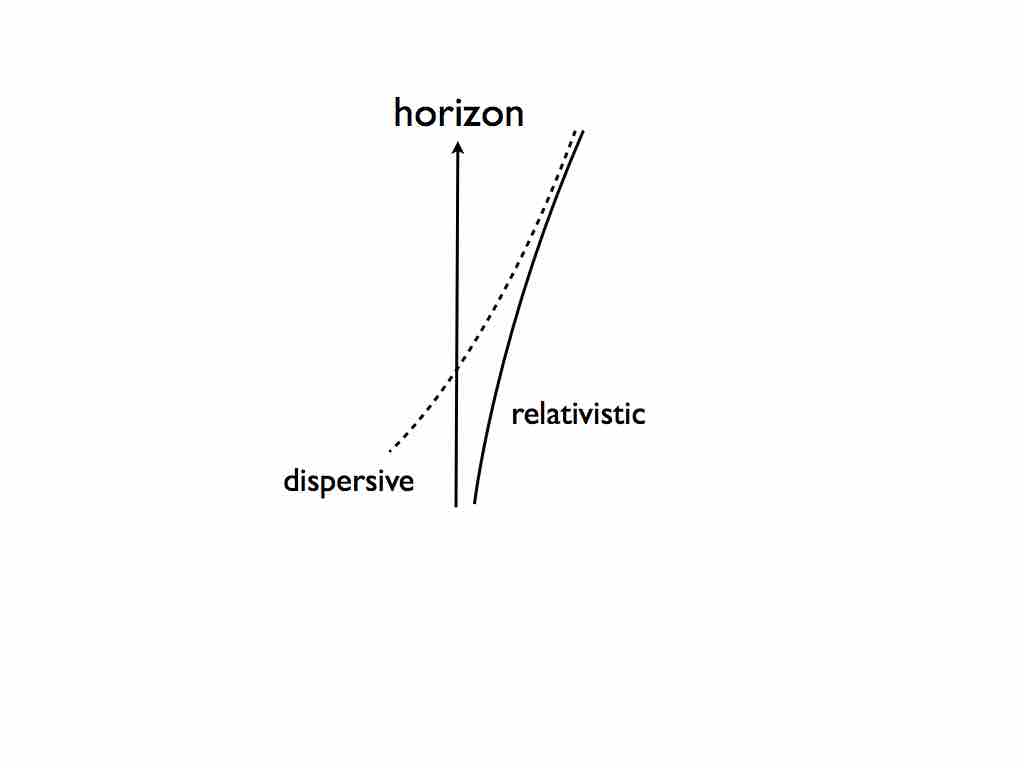}
\caption{Dispersion (dashed line) eliminates the piling up of the modes on the horizon (solid line).}
\label{disp-lin}
\end{figure}

\noindent This is the key feature that a BH and a supersonically flowing BEC have in common: the presence in the spectrum of negative energy
states. These lay inside the horizon for the BH (where the Killing vector associated to stationarity is spacelike) and in the supersonic region for the BEC. As a consequence, in both systems one can have production of particles out of the vacuum without violating energy conservation. Particles are produced in entangled pairs, one with positive energy and the other (the partner) with negative energy. This is how vacuum fluctuations are converted in real on shell particles (phonons in the BEC case).
So without using the gravitational analogy the occurrence of Hawking like radiation in BEC can be firmly established starting from the Bogoliubov theory eqs. (\ref{sei}), (\ref{sette}).
One can then calculate, given an acoustic BH like profile for the flow, the spectrum of phonons emitted in the subsonic region of the BEC. In terms
of S-matrix theory the process corresponds to the conversion of a negative energy $k_4^{(-)}$ incoming mode in the supersonic
region to a positive energy $k_u^{(+)}$ outgoing mode in the subsonic one. 
There is no transplanckian problem since the dispersive character of eq. (\ref{diciotto}) eliminates the piling up of the modes on the horizon characteristic of the relativistic dispersion relation (\ref{diciannove}) (see Fig. \ref{disp-lin}) responsible of the infinite blueshift of the frequency.

\noindent The obtained spectrum can then be compared to the one expected on the basis of the analogy, namely $N_\omega$ of eq. (\ref{uno})
with a temperature given by eq. (\ref{sedici}) \cite{ma-pa, fi-pa}. 

\noindent First of all one should note that the emission is cut-off at $\omega_{max}$. For $\omega>\omega_{max}$ there are no more negative energy states (see Fig. \ref{bec-sup}). Without the presence of a partner a Hawking quantum can not materialize out of the vacuum.
This is completely missed by the analogy.  Given this, one finds that for $\xi\ll \frac{c_s}{\kappa}$ the emission is thermal at exactly the Hawking temperature (\ref{sedici}). So in this regime the predictions of the analogy are reliable. On the other hand for 
$\xi\gg \frac{c_s}{\kappa}$ the emission at low $\omega$ is still thermal but the temperature differs from the one associated to the surface gravity.
Furthermore, significant deviations from the prediction of the analogy emerge in limiting cases. 

\section{Limiting cases}

\begin{figure}[h]
\centering \includegraphics[angle=0, height=2.4in] {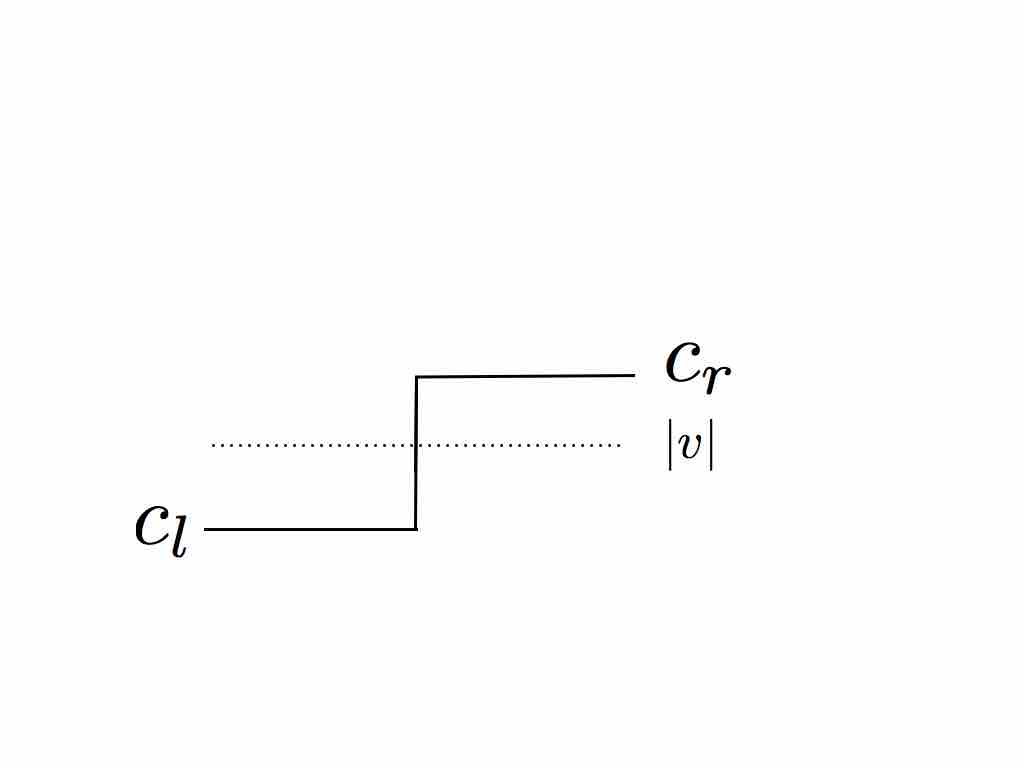}
\caption{Step-like profile for an acoustic black hole.}
\label{step}
\end{figure}

Consider the following step-like profile for the BEC flow (Fig. \ref{step}) for which the velocity $v$ is constant and the sound velocity $c_s$ equals $c_r>|v|$ in the subsonic region while equals $c_l<|v|$ in the supersonic region. 

\noindent One finds at low $\omega$ an emission temperature proportional to $\xi^{-1}$ \cite{rpc, lrcp}
\be
T_{step}=\frac{\hbar}{k_B}\frac{(c_r+v)(v^2-c_r^2)}{(c_r-v)(c_r^2-c_l^2)}\frac{2c_r}{c_l\xi}\ .
\ee

$T_{step}$  can be considered as the maximum temperature achievable for monotonic profiles with fixed asymptotic $c_r$ and $c_l$ . If the step like profile is obtained by a limiting procedure of smooth profiles, like for example  $\tanh(x/L)$ with $L\to 0$, the analogy would predict a continuously increasing surface gravity and hence of $T_H$, diverging in the step limit.The healing length, which characterizes the dispersive regime of (\ref{diciotto}), provides therefore an upper cutoff for the emission temperature. So the analogy in this case is completely unreliable . 

\begin{figure}[h]
\centering \includegraphics[angle=0, height=2.4in] {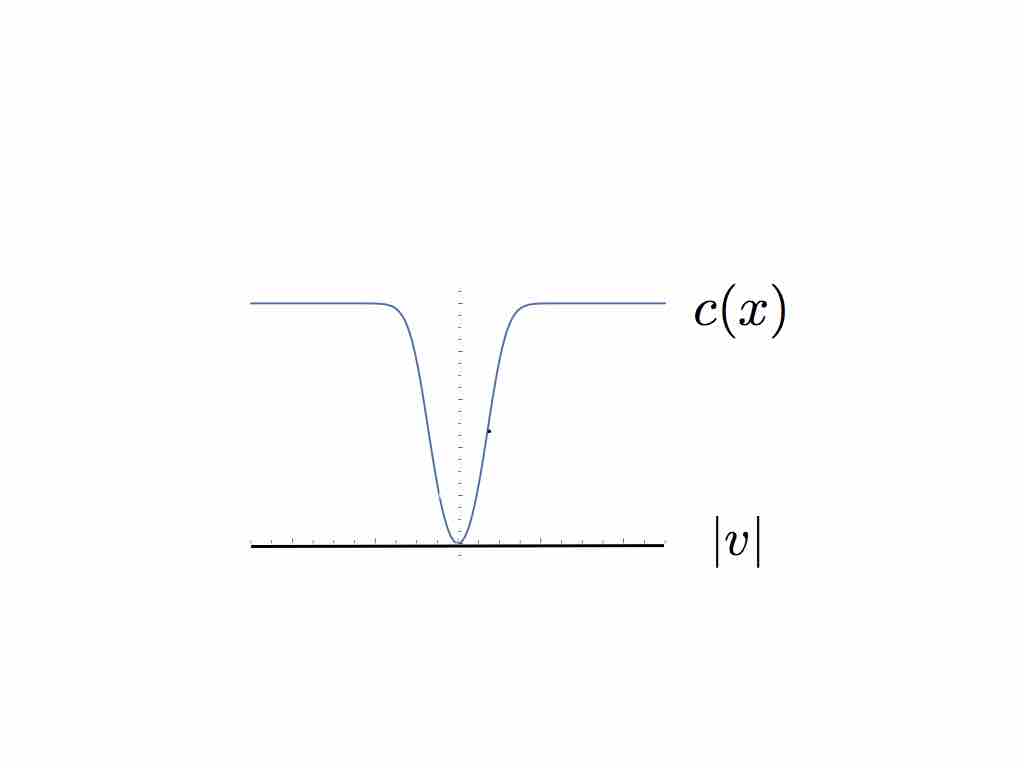}
\caption{Zero temperature flow profile with no supersonic region.}
\label{zeroT}
\end{figure}

 Consider now the opposite situation in which $\kappa=0$. The usual example in the gravitational case would be the extreme
Reissner-Nordstr\"om BH described, in Schwarzschild coordinates, by
\be \label{ventuno}
ds^2=-(1-\frac{GM}{c^2r})^2dt^2 + \frac{dr^2}{(1-\frac{GM}{c^2r})^2}+r^2d\Omega^2
\ee
for which $\kappa=0$ and $T_H=0$, i.e. no Hawking emission.
A corresponding BEC flow profile would be something like the one depicted in Fig. \ref{zeroT}.

\noindent With this profile there is no phonons emission neither for the BEC. The reason is that there is no supersonic region and so there are no 
negative energy states. Similarly for the BH, the Killing horizon associated to stationarity is everywhere timelike (it becomes null just on the horizon $r=\frac{GM}{c^2}$). 

\begin{figure}[h]
\centering \includegraphics[angle=0, height=2.4in] {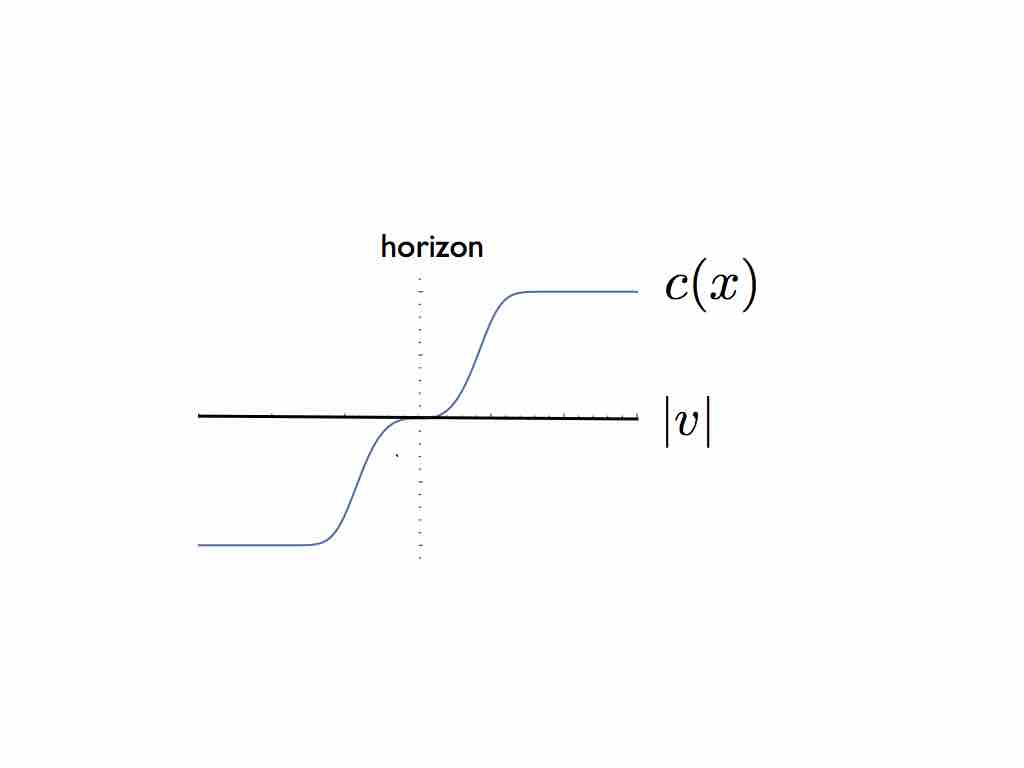}
\caption{Zero temperature flow profile with a supersonic region.}
\label{cub}
\end{figure}

 More tricky is the following example.
Consider a BEC with a profile like the one depicted in Fig. \ref{cub}, for example $c(x)\propto \tanh(x^3)$.

\noindent A corresponding BH metric would be
\be \label{ventidue}
ds^2=-g_{00}dt^2+ g_{00}^{-1}dr^2+r^2d\Omega^2\ ,
\ee
where $g_{00}\sim (r-r_H)^3$ for $r$ approaching the horizon $r_H$. Having $\kappa=0$ this BH does not emit and the analogy
would predict the same for the BEC. This is not the case since for the BEC, using the Bogoliubov theory, one finds an emission characterized by $N_\omega \sim const.$ at low $\omega$  \cite{finazzi}. 
The presence of a supersonic region with negative energy states available guarantees the emission in the BEC. 
The problem is to 
understand why the BH described by the metric (\ref{ventidue}) does not emit even having, unlike the extreme Reissner-Nordstr\"om BH of eq. (\ref{ventuno}), an interior region where negative energy states exist (the Killing vector is spacelike there). 

In general, for BHs the emission is triggered by the collapse of the star across the horizon. 
There is an initial transient emission which depends on the details of the collapsing star which rapidly decays in time leaving at late time a steady thermal flux independent on the collapse details being a function only of the surface gravity $\kappa$ of the 
resulting BH horizon. This latter is Hawking radiation, described by eqs. (\ref{uno}), (\ref{due}). 
The presence of this steady thermal emission then requires a BH region and a nonvanishing $\kappa$. When $\kappa=0$
there is just a transient emission which rapidly decreases to zero and no steady thermal flux. 
To get a steady emission in the BEC case it is not necessary to take into account the formation of the sonic horizon.
Given the dispersive character of the dispersion relation eq. (\ref{diciotto}) one has emission even in a stationary flow simply because
of the existence of the $k_4^{(-)}$ mode capable of propagating upstream in the supersonic region as we have seen.

\section{Low energy-emission and gray-body factor}

The thermality of Hawking radiation is not completely exact because of the backscattering of the modes caused by the curvature of the space-time, and the expression (\ref{uno}) of the emitted particles gets a correction, namely
\be \label{ventitre}
N_\omega=\frac{\Gamma(\omega)}{e^{\frac{\hbar\omega}{k_BT_H}}-1}\ ,
\ee
where $\Gamma(\omega)$ is the so called ``gray-body'' factor which accounts for the backscattering (see for instance \cite{dewitt}). $\Gamma(\omega)$
is the probability for an outgoing mode created near the horizon to reach infinity. 
For asymptotically flat BHs \cite{page}
\be \Gamma(\omega)\sim A_H\omega^2\ ,
\ee
where $A_H$ is the area of the BH horizon. For $\omega\to 0$ outgoing modes are almost completely reflected back to the horizon.
This correction makes $N_\omega$ of eq. (\ref{ventitre}) infrared finite.

\noindent In the BEC case  the small $\omega$ behaviour can be well analyzed by the hydrodynamical limit of the BdG equations namely eqs. (\ref{quattordici}), (\ref{quindici}) and one finds a quite different result : $\Gamma(\omega)$ goes to a constant
at small $\omega$ \cite{abfp, afb}
\be \label{venticinque}
\Gamma(\omega)= \frac{c_r v}{(v+c_r)^2}\ .
\ee
The reason is that in the BEC the effective potential responsible for the backscattering of the modes allows for the existence of an everywhere bounded solution of the $\omega\to 0$ limit of the wave equation. In the BH case the corresponding limit gives a solution 
growing like $r$ at infinity. So the graybody factor of eq. (\ref{venticinque}) does not cancel the $\frac{1}{\omega}$ divergence coming from the Planckian distribution: the emission of the BEC black hole is dominated by soft phonons.

\section{Correlations}

\begin{figure}[h]
\centering \includegraphics[angle=0, height=2.6in] {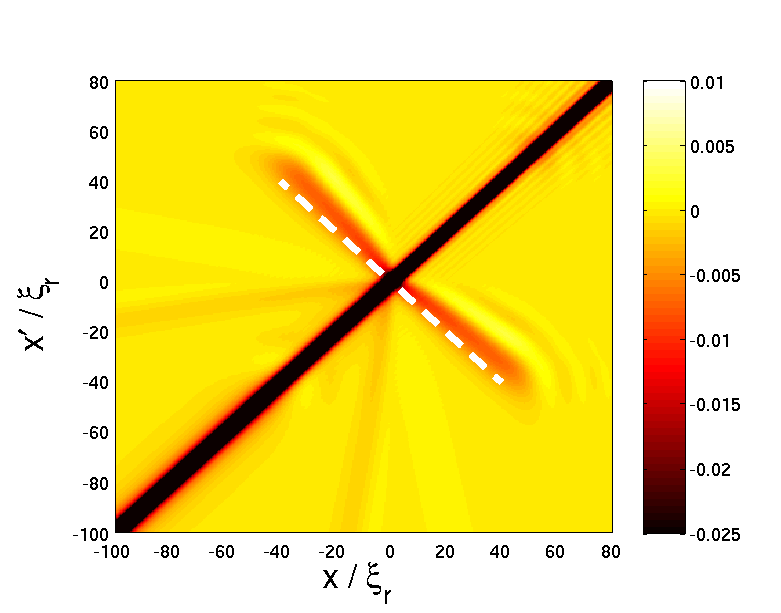}
\caption{Hawking quanta-partner peak in the density correlator (the dotted line is the hydrodynamical prediction).}
\label{corr}
\end{figure}

Having established the solid foundations of the existence of Hawking like radiation in BEC one hopes this radiation to be revealable 
in laboratory experiments. The major problem one has to overcome is that in the most favourable case the emission temperature
expected is at least one order of magnitude lower than the background temperature of the BEC ($10^{-9}K$). 
So thermal phonons  can completely cover the Hawking signal one is searching for. 
However it was realized that, being Hawking radiation basically a pair production process, one can look for the 
correlation between the pair members, i.e. the Hawking particle and the partner inside the hole. The unambiguous signature of this 
correlation is a peak in the correlation functions, for example the density-density. It was predicted in \cite{bffr}, using the hydrodynamic approximation of the BEC theory, that thermal analog Hawking radiation gives rise to the (one-time) nonlocal correlator 
\be \label{hpcorr}
\langle \hat n_1 (t,x)\hat n_1(t,x) \rangle \sim - \frac{\kappa^2}{\cosh^2(\frac{\kappa}{2}(\frac{x}{c_r-|v|}+\frac{x'}{|v|-c_l}))},
\ee
for points located on the both sides and sufficiently away from the horizon where the profile is to a good approximation homogeneous.
We see that that height and width of this correlator is fixed by the black hole surface gravity $\kappa$, and disappears when there is no emission.
Such a profile has a stationary peak along the line \be \label{peak} \frac{x}{c_r-|v|}+\frac{x'}{|v|-c_l}=0, \ee
and this in turns corresponds to a steady emission of phonons by the horizon ($x=0$), one of which propagating upstream ($k_u^{(+)}$) in the region exterior to the horizon with velocity $c_r-|v|$ and the other (the partner, $k_u^{(-)}$) being dragged by the flow in the supersonic region and propagating  with velocity $c_l-|v|$. It  was then shown that this signature persists in the full BEC theory \cite{cfrbf},
and, importantly, that it is still there in the presence of a thermal bath at a temperature $T>T_H$. 
See Fig. \ref{corr}. 
The peak at (\ref{peak}) is the smoking gun of the Hawking 
effect in BECs. 

\noindent In a series of experiments, starting in 2016, J. Steinhauer \cite{jeff} was indeed able to reveal this peak giving the first experimental evidence of Hawking like radiation in BECs (see also \cite{jeffetal}). 
Finally, the quantum signature of Hawking radiation (i.e. the fact that the emitted pairs come from the vacuum, i.e. they are entangled, rather than the thermal background) is better characterized in terms of the Fourier transform of the density correlator  \cite{bu-pa, jefft}, or the momentum correlator \cite{mom-corr1, mom-corr2}.

\bigskip\bigskip\bigskip
\noindent {\bf{Acknowledgments:}}  
A.F. acknowledges partial financial support by the Spanish Grants PID2020-116567GB-C21, PID2023-149560NB-C21 funded by MCIN/AEI/10.13039/501100011033, and by the Severo Ochoa Excellence Grant CEX2023-001292-S.

{}

\end{document}